\documentclass[%
 reprint,
 amsmath,amssymb,
 aps,
float fix,
]{revtex4-2}
\usepackage{graphicx}
\usepackage{dcolumn}
\usepackage{bm}
\usepackage{amsmath}

\allowdisplaybreaks[4]
\usepackage{subfigure}
\usepackage{xcolor}
\usepackage{soul}
\soulregister\cite7
\soulregister\ref7

\usepackage{amsmath}

\usepackage[colorlinks,linkcolor=blue,citecolor=blue,urlcolor=blue,hyperindex,breaklinks]{hyperref}
\allowdisplaybreaks[4]
\usepackage{subfigure}

\begin{document}
\title{Quantum heat valve and diode of strongly coupled defects in amorphous material}
\author{Yu-qiang Liu, Yi-Jia Yang, Ting-ting Ma, Zheng Liu, and Chang-shui Yu}
\email{Electronic address: ycs@dlut.edu.cn}
\affiliation{$^1$School of Physics, Dalian University of Technology, Dalian 116024,
P.R. China}

\date{\today}

\begin{abstract}
The mechanical strain can control the frequency of two-level atoms in amorphous material. In this work, we would like to employ two coupled two-level atoms to manipulate the magnitude and direction of heat transport by controlling mechanical strain to realize the function of a thermal switch and valve. It is found that a high-performance heat diode can be realized in the wide Piezo voltage range at different temperatures.  We also discuss the dependence of the rectification factor on temperatures and couplings of heat reservoirs. We find that the higher temperature differences correspond to the larger rectification effect. The asymmetry system-reservoir coupling strength can enhance the magnitude of heat transfer, and the impact of asymmetric and symmetric coupling strength on the performance of the heat diode is complementary. It may provide an efficient way to modulate and control heat transport's magnitude and flow preference. This work may give insight into designing and tuning quantum heat machines.
\end{abstract}
\pacs{03.65.Ta, 03.67.-a, 05.30.-d, 05.70.-a}
\maketitle
\section{Introduction}

The theoretical and experimental advances of quantum heat transport, including heat conductance and control by external fields, lead to the rapid development of quantum thermodynamics \cite{pekola2021colloquium}. Moreover, recent advances in the superconducting quantum circuit \cite{you2011atomic, devoret2013superconducting, PhysRevB.91.094517} based on flexibility and scalability have provided new insight into quantum heat transport \cite{pekola2021colloquium, gubaydullin2022photonic} and quantum thermodynamics \cite{vinjanampathy2016quantum}. Besides, manipulating electronic heat current in solid-state thermal circuits also provides a reference for the control heat current in quantum systems \cite{martinez2015rectification}. The manipulation and control of heat current at the quantum level as a basic research field of quantum thermodynamics have been significant progress. So far, various forms of quantum thermal machines have been proposed and designed, such as refrigerators and heat engines \cite{PhysRevLett.105.130401, PhysRevLett.108.070604, PhysRevE.90.052142, brask2015autonomous, PhysRevB.94.235420, PhysRevE.105.034112, PhysRevApplied.19.034023}, rectifier \cite{tesser2022heat, PhysRevE.95.022128, PhysRevE.89.062109, PhysRevE.90.042142, PhysRevLett.120.200603, PhysRevE.99.032136, PhysRevA.105.052605, PhysRevB.79.144306, PhysRevB.103.104304, PhysRevB.101.075417, PhysRevE.99.042121, PhysRevE.96.012122, PhysRevApplied.15.054050, PhysRevE.107.044121}, transistor \cite{PhysRevLett.116.200601, PhysRevE.98.022118, PhysRevE.106.024110, PhysRevB.101.184510, PhysRevE.99.032112, PhysRevB.99.035129, PhysRevA.97.052112}, thermometer \cite{PhysRevLett.119.090603, PhysRevB.105.235412}, valve \cite{ronzani2018tunable, liu2023quantum}, and attracted extensive attention from plenty of researchers. As a vital part of thermal machines, thermal rectification devices mainly refer to diodes and transistors. As a two-terminal device, a quantum thermal diode can induce heat flows inclined in one direction.   As a three-terminal device, quantum thermal transistors can realize that a weak heat current can amplify the heat currents of the two other terminals. Quantum thermal diodes and transistors are analogous to their electronic counterpart and have been designed in plenty of systems such as atoms \cite{PhysRevLett.116.200601, PhysRevE.95.022128, liu2021common, PhysRevE.99.032112, PhysRevE.98.022118, PhysRevE.89.062109, liu2021common, PhysRevE.107.064125}, quantum circuits \cite{PhysRevB.79.144306, PhysRevB.101.184510, PhysRevE.107.044121}, spin chains \cite{PhysRevE.90.042142, PhysRevE.99.032136, PhysRevLett.120.200603, PhysRevA.105.052605}, and quantum dots \cite{tesser2022heat, PhysRevB.101.075417, PhysRevB.99.035129}. These proposals mainly focus on the modulation of the temperature of the reservoirs. A natural question arises as to whether or not other proposals exist to realize the manipulation of heat transport.

 \begin{figure}[!htbp]
 \subfigure{
\centering \includegraphics[width=0.95\columnwidth]{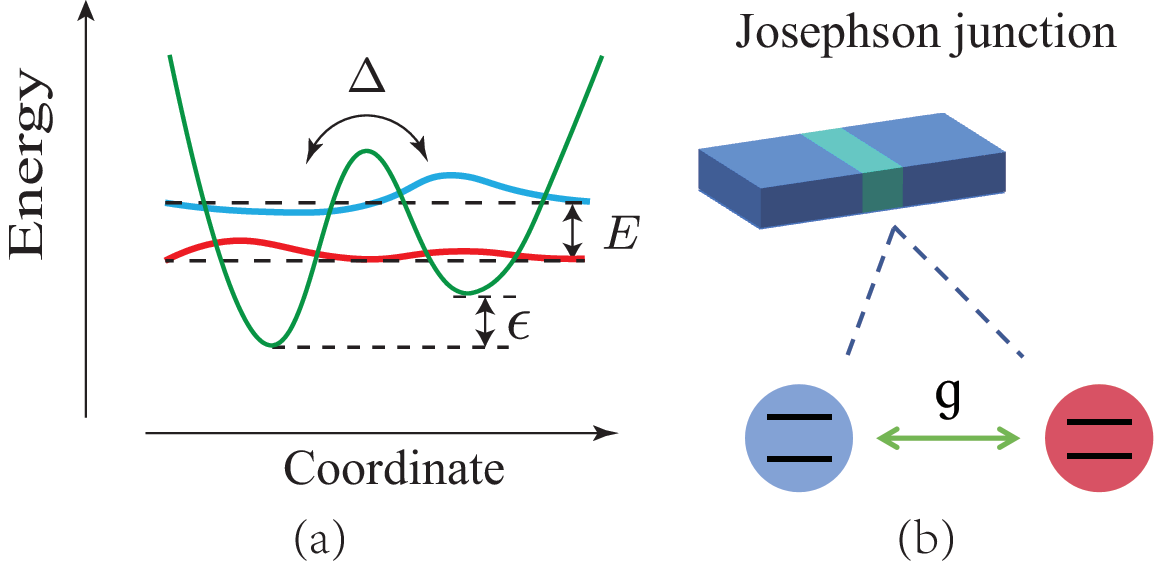} }
\subfigure{
\centering \includegraphics[width=0.95\columnwidth]{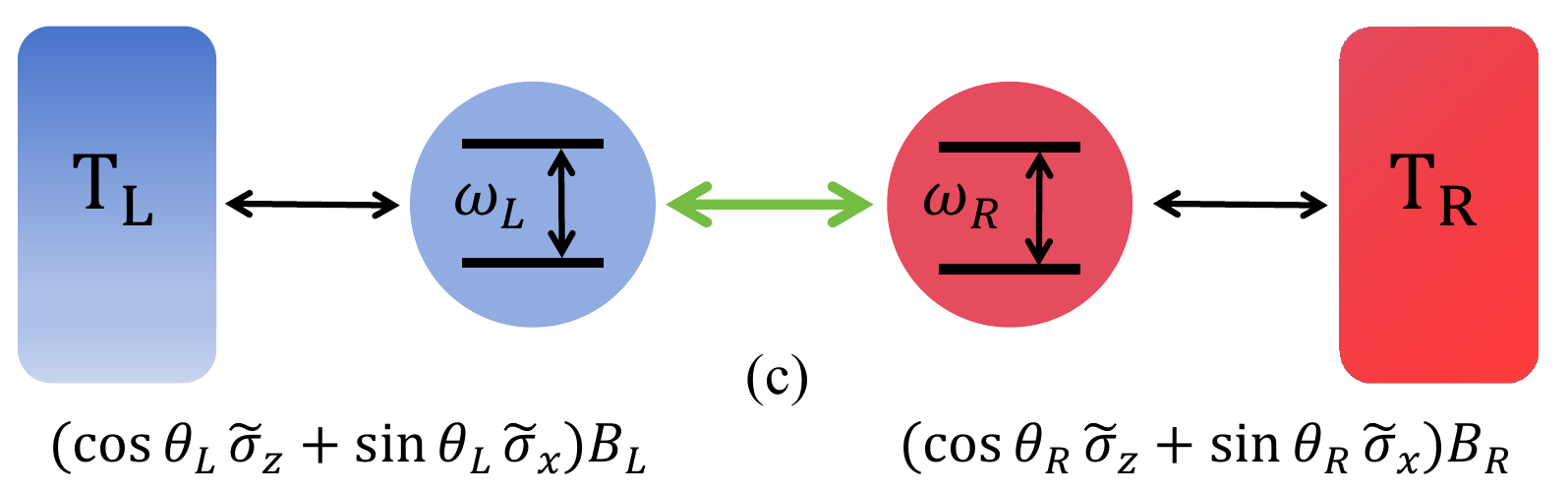} }\caption{(a) The double-well potential system in the two-level atom limit. $\Delta$, $\epsilon$ denote the tunneling energy and asymmetry energy and $E$ is the level splitting with $E=\sqrt{\Delta^2+\epsilon^2}$. (b) The model consists of two coherent coupling two-level atoms embedding in the tunnel barrier of a Josephson junction (JJ). (c) Schematic diagram of quantum heat diode and valve. Intuitively, the mechanism of asymmetry of heat transport can be roughly understood in two contributions: the off-resonantly interacting atoms and the different system-reservoir couplings.  }
\label{two_level_atom} 
\end{figure}

In recent years, some researchers have tried to explore other modulation forms of heat transport, such as period driving and external magnetic flux. Nikhil Gupt et al. employ periodic control to realize a quantum thermal transistor \cite{PhysRevE.106.024110}. Pedro Portugal et al. research heat transport using a two-level system via a periodically modulated
temperature \cite{PhysRevB.104.205420}. B. Karimi et al. \cite{karimi2017coupled} employ two superconducting qubits to design a flux-modulation heat switch. 
Alberto Ronzani et al. realize a heat valve with a transmon qubit by modulating the applied magnetic flux \cite{ronzani2018tunable}. Meng Xu et al. discuss heat transfer with an external magnetic flux tunable transmon qubit \cite{PhysRevB.103.104304}. Jorden Senior et al. design a magnetic flux-controlled heat diode \cite{senior2020heat}. 
In this context, researching different modulation ways of heat current has gradually become a promising topic. Recently, Lisenfeld et al. first directly and experimentally observed two strongly interacting and coherent two-level systems embedding in the tunnel barrier of a Josephson junction (JJ) by modulating the mechanical strain \cite{lisenfeld2015observation}. Also, a two-level system or small group of atoms in an amorphous material is one of the most promising models \cite{muller2019towards, grabovskij2012strain}. In addition, the two-level defect can also be coupled with the optomechanical system to achieve phonon blockade and preparation of non-classical states. \cite{PhysRevLett.110.193602}. Given that the changes of two-level atom frequencies depend on the applied piezo voltage, it is advantageous to employ the piezo voltage to control the heat transport for realizing some function. Inspired by this, we will use this coupled defect system to design a mechanical strain-controlled heat valve and diode in amorphous solids.

This work considers the coupling defects to design a perfect heat diode and controllable heat valve. We first show that a well-performance heat valve and diode can be realized by modulating the Piezo voltage, and the maximal rectification can be obtained when considering the resonant case. We find that a large rectification effect can be discovered at resonant frequencies with a large temperature bias. Next, we investigate how the asymmetric system-reservoir coupling 
strength further influences the performance of the heat valve and diode. We find that the stronger asymmetric coupling strength can realize the larger heat transport, and the performance of heat diodes employing asymmetric and symmetric couplings can be complementary. Besides, we also discuss the dependence of the performance of a heat diode on temperature and find that the higher temperature difference allows for greater heat current rectification.

The structure of this paper is organized as follows.  In Sec. \ref{Sec. II}, we introduce the Hamiltonian of two coupled two-level atoms and their dissipation and derive the global master equation. In Sec. \ref{Results and discussions}, we discuss a modulated heat valve and the realization parameters range of perfect rectification. The conclusion is given in Sec. \ref{conclusion}.

\section{Physical model and dynamics} \label{Sec. II}

Josephson junctions consist of aluminum, and its oxide has become a common functional element in the design of superconducting circuits \cite{PhysRevB.97.180505}. In general, the disordered thin film AlOx of the junction exists in atomic tunneling systems, and it is described by a double-well system in the two-state limit, as shown in Fig. \ref{two_level_atom} (a). We follow the model consisting of two coherent coupling two-level atoms embedding in the tunnel barrier of a Josephson junction (JJ) in Fig. \ref{two_level_atom} (b), and the details can refer to Refs. \cite{lisenfeld2015observation, de2022strain}. 
The Hamiltonian of the coupling two atoms can read as \cite{lisenfeld2015observation, de2022strain} ($\hbar=k_{B}=1$) 
\begin{align} \label{ori H_S}
H_S=\frac{1}{2} \sum_{\mu=L, R} [\varepsilon_{\mu} \sigma^{z}_{\mu}+ \Delta_{\mu} \sigma^{x}_{\mu}]+\frac{1}{2} g \sigma^{z}_{L} \sigma^{z}_{R},
\end{align}
where $\sigma^{z}_{\mu}$, $\sigma^{x}_{\mu}$ are Pauli spin operators, $\Delta_{\mu}$ denote the tunneling energy, $\varepsilon_{\mu}$ are the asymmetry energy, and $g$ is the coupling strength of two atoms. Experimentally, one can slightly bend the sample chip by employing a piezo actuator to manipulate and control the asymmetry energy in-situ $\varepsilon_{\mu}=\varepsilon_{\mu}(V_p)=c_{\mu}(V_{p}-V_{0 \mu})$ \cite{muller2019towards, lisenfeld2016decoherence}.
A unitary transformation can lead to this equation 
\begin{equation} \label{transformation}
\sigma^{z}_{\mu}\rightarrow \cos \theta_{\mu} \tilde{\sigma}_{\mu}^{z}+\sin \theta_{\mu} \tilde{\sigma}_{\mu}^{x},
\end{equation}
and the corresponding system Hamiltonian $(\ref{ori H_S})$ can also be transformed as 
\begin{align} \label{system Hamiltonian}
\tilde{H}_S=\frac{\omega_L}{2} \tilde{\sigma}_L^{z}+\frac{\omega_R}{2} \tilde{\sigma}_R^{z}+\frac{g_{\|}}{2} \tilde{\sigma}_L^{z} \tilde{\sigma}_R^{z}+\frac{g_{\perp}}{2} \tilde{\sigma}_L^{x} \tilde{\sigma}_R^{x},
\end{align}
where $\omega_{\mu}=\sqrt{\varepsilon_\mu^2\left(V_p\right)+\Delta_{\mu}^2}, g_{\|}=g \cos \theta_{L} \cos \theta_{R}$ and $g_{\perp}=g \sin \theta_{L} \sin \theta_R$ with $\cos \theta_{\mu}=\varepsilon_{\mu}/\omega_{\mu}$ denote transversal and longitudinal coupling, respectively and they are easily identifiable in experiment \cite{lisenfeld2015observation} as well as two minor energy shifts terms can be neglected \cite{lisenfeld2015observation, de2022strain}.
One can diagonalize the system Hamiltonian (\ref{system Hamiltonian}), and the eigenvalues can read as
\begin{equation}
\begin{split}
& \epsilon_{1,2}=\mp \frac{1}{2} \sqrt{\left(\omega_L+\omega_R\right)^2+g_{\perp}^2}+\frac{g_{\|}}{2}, \\
& \epsilon_{3,4}= \pm \frac{1}{2} \sqrt{\left(\omega_L-\omega_R\right)^2+g_{\perp}^2}-\frac{g_{\|}}{2}.
\end{split}
\end{equation}
We can express it as a vector form $\vert\epsilon \rangle=[\epsilon_{1}, \epsilon_{2}, \epsilon_{3}, \epsilon_{4}]$,
and the corresponding eigenstates are
\begin{equation}
\begin{split}
& \left|\epsilon_1\right\rangle=\cos (\alpha / 2)|g, g\rangle-\sin (\alpha / 2)|e, e\rangle, \\
& \left|\epsilon_2\right\rangle=\cos (\alpha / 2)|e, e\rangle+\sin (\alpha / 2)|g, g\rangle, \\
& \left|\epsilon_3\right\rangle=\cos (\beta / 2)|e, g\rangle+\sin (\beta / 2)|g, e\rangle, \\
& \left|\epsilon_4\right\rangle=\cos (\beta / 2)|g, e\rangle-\sin (\beta / 2)|e, g\rangle.
\end{split}
\end{equation}
with $\tan \alpha=\frac{g_{\perp}}{\omega_{L}+\omega_{R}}$, $\tan \beta=\frac{g_{\perp}}{\omega_{L}-\omega_{R}}$.

 \begin{figure}[!htbp]
\centering
\subfigure[]{
 \includegraphics[width=0.42\columnwidth]{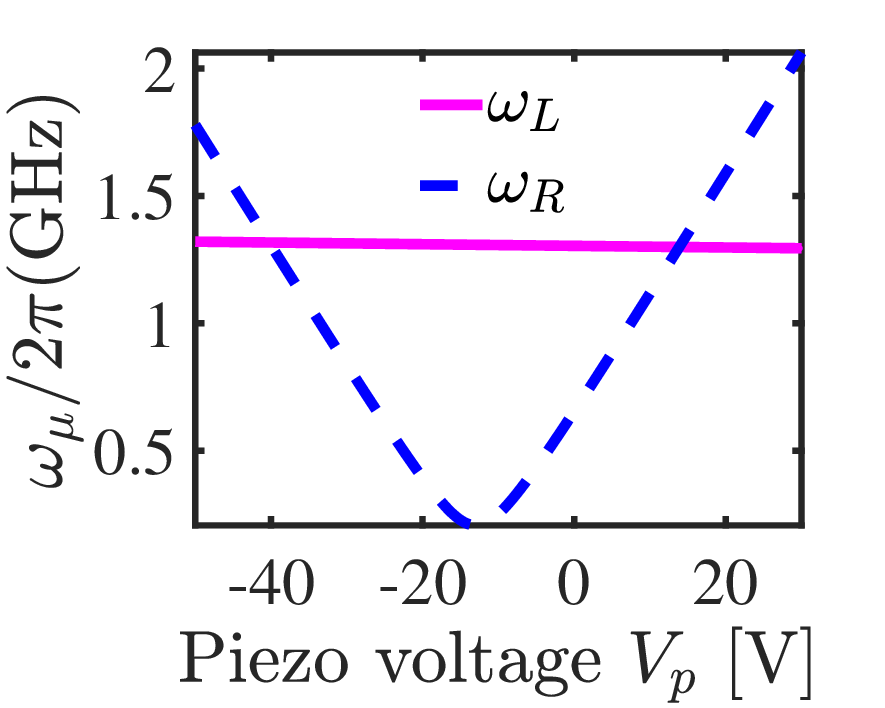
} }\hspace{-0.2cm}
\centering
\subfigure[]{
 \includegraphics[width=0.48\columnwidth]{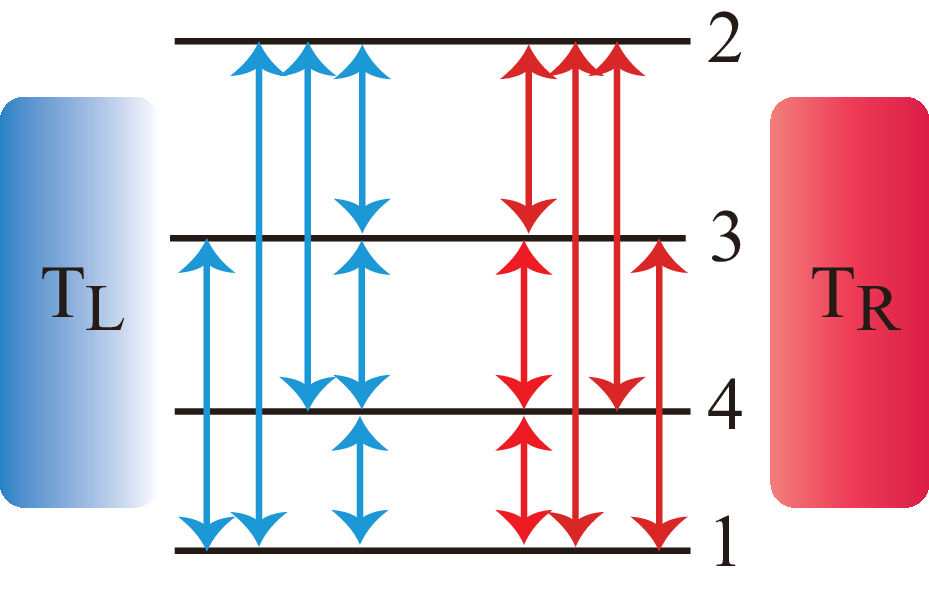
} } 
\caption{(a), The Bohr frequencies $\omega_{\mu}$ versus the Piezo voltage $V_P$; (b), transition channels of open system. System parameters: the tunneling energy $\Delta_L=7.5 \mathrm{GHz}$ and $\Delta_R=1.3 \mathrm{GHz}$; the asymmetry energy $\varepsilon_L=c_L V_p-3.3[\mathrm{GHz}]$ and $\varepsilon_R\left(V_p\right)=c_R\left(V_p+13[\mathrm{V}]\right)$ ; $c_L=5 \mathrm{MHz} \mathrm{V}^{-1}$ and $c_R=0.3 \mathrm{GHz} \mathrm{V}^{-1}$, the coupling strength $g=850 \mathrm{MHz}$.}
\label{Fig_frequency} 
\end{figure}

 The system inevitably dissipates the energy to the external environment. We let the two atoms coupled to an independent reservoir. In general, the environment can be modeled by a collection of independent harmonic oscillators, and the free Hamiltonian can read as \cite{RevModPhys.59.1} 
\begin{equation}
\tilde{H}_{R}=\sum_{\mu}\omega_{\mu k} c^{\dagger}_{\mu k} c_{\mu k},
\end{equation}
where $\omega_{\mu k}$ denotes the frequency of reservoir $\mu$ at mode $k$ th, and $c_{\mu k}$, $c^{\dagger}_{\mu k}$ are bosonic annihilation and
creation operators of the reservoir modes. The reservoir can be modeled by the normal-metal resistor \cite{senior2020heat} or a general noiseless resistor with a
fluctuating voltage source \cite{RevModPhys.82.1155, https://doi.org/10.1002/qute.202100054}.   
We can follow a Caldeira-Leggett-type
system-bath Hamiltonian \cite{PhysRevB.30.1208, PhysRevB.77.041303, PhysRevB.81.144510, PhysRevB.80.174103}
\begin{align} \label{s-r coupling}
\tilde{H}_{S-R}= B_{\mathrm{L}}\left(\sigma_L^{z}\otimes 1\right)+ B_{\mathrm{R}}\left(1 \otimes \sigma_R^{z}\right),
\end{align}
where $B_{\mathrm{\mu}}=\sum_{k}\kappa_{\mu k}(c_{\mu k}+c^{\dagger}_{\mu k})$ and $\kappa_{\mu k}$ represents the coupling strength of atom-reservoir. The coupling of Eq. (\ref{s-r coupling}) is described by a spectral density \cite{RevModPhys.89.015001} 
\begin{equation}
J(\omega)=\sum_{k} \kappa_{k}^2 \delta(\omega-\omega_{k}).
\end{equation} 
Hence, one can obtain the full Hamiltonian of system, environment, and their interaction,
\begin{equation} \label{total Hamiltonian}
H=\tilde{H}_S+\tilde{H}_R+\tilde{H}_{S-R}.
\end{equation} 
With transformation (\ref{transformation}), the system-reservoir interaction Hamiltonian (\ref{s-r coupling}) can rewrite as 
\begin{equation} \label{transformed s-r coupling}
H_{S-R}=B_{\mathrm{L}}\left(S_L\otimes 1\right)+B_{\mathrm{R}}\left(1 \otimes S_R\right),
\end{equation}
with transformed jump operators $S_{\mu}=\cos \theta_{\mu} \tilde{\sigma}_{\mu}^{z}+\sin \theta_{\mu} \tilde{\sigma}_{\mu}^{x}$. In $H_S$ representation, the total Hamiltonian (\ref{total Hamiltonian}) can be rewritten as 
\begin{equation}
H=\sum^{4}_{j=1}\left|\epsilon_j\right\rangle\left\langle \epsilon_j\right| +\tilde{H}_{R}+H^{\prime}_{S-R},
\end{equation}
and
\begin{equation}
H_{S-R}^{\prime}=\sum_{\mu, k, l} \kappa_{\mu k}\left[c_{\mu k}^{\dagger} S_{\mu l}\left(\omega_{\mu l}\right)+c_{\mu k} S_{\mu l}^{\dagger}\left(\omega_{\mu l}\right)\right].
\end{equation}
Here the eigenoperators $S_{\mu l}$ satisfy the commutation relation, $[H_{S}, S_{\mu l}]=-\omega_{\mu l} S_{\mu l}$, and $\omega_{\mu l}$ are the corresponding eigenfrequencies with subscript $l=0, 1, 2, 3, 4, 5, 6$ shown in Appendix \ref{Appendix A}. Note that the values of $\omega_{\mu l}$ depends on the different transition channels $\left|\epsilon_{j} \right\rangle  \leftrightarrow \left|\epsilon_{m} \right\rangle$ .

 To study the dynamics of the system, we follow the stand method to derive the master equation based on Born-Markov-secular approximation, and the Lindblad form in the Schrödinger picture can be written as \cite{breuer2002theory}
\begin{align} \label{master equation}
\dot{\rho}=-i\left[H_S, \rho\right]+\mathcal{L}_L[\rho]+\mathcal{L}_R[\rho],
\end{align}
where the dissipator $\mathcal{L}_\mu[\rho]$ is given by
\begin{align}
\nonumber
\mathcal{L}_\mu[\rho]&=  \sum_{l=1}^{6} J_\mu\left(-\omega_{\mu l}\right)\left[2 S_{\mu l}\left(\omega_{\mu l}\right) \rho S_{\mu l}^{\dagger}\left(\omega_{\mu l}\right)\right. \\ \nonumber
& \left.-\left\{S_{\mu l}^{\dagger}\left(\omega_{\mu l}\right) S_{\mu l}\left(\omega_{\mu l}\right), \rho\right\}\right] \\ \nonumber
& +J_\mu\left(\omega_{\mu l}\right)\left[2 S_{\mu l}^{\dagger}\left(\omega_{\mu l}\right) \rho S_{\mu l}\left(\omega_{\mu l}\right)\right. \\
& \left.-\left\{S_{\mu l}\left(\omega_{\mu l}\right) S_{\mu l}^{\dagger}\left(\omega_{\mu l}\right), \rho\right\}\right].
\end{align}
where the spectral densities can take $J_\mu\left(\pm\omega_{\mu l}\right)=\gamma_{\mu}(\omega_{\mu l}) n_{\mu}(\pm\omega_{\mu l})$ with $\gamma_{\mu} (\omega)=\sum_{k} |\kappa_{\mu k}|^2 \delta(\omega-\omega_{\mu k})$ \cite{PhysRevE.85.061126}, and $n_{\mu}(\omega)=\frac{1}{\exp(\frac{\omega}{T_{\mu}})-1}$ is a Bose-Einstein function. We assume the dissipation rates are independent of frequencies for simplification, i.e., $\gamma_{\mu}(\omega_{\mu l})=\gamma_{\mu}$. In the system representation, the master equation (\ref{master equation}) is divided into diagonal and off-diagonal entries of the density operator. At steady-state, the only diagonal elements of the density matrix, termed populations \cite{schaller2014open, PhysRevE.107.044121} can remain, and the differential equation for the diagonal entries $\rho_{jj}$ read as  
\begin{equation}
\dot{\rho}_{jj}=\underset{\mu}{\sum}\underset{m}{\sum}\Gamma_{jm}^{\mu}(\rho)\text{,}\label{differential equation}
\end{equation}
where 
\begin{equation} \nonumber
\Gamma_{jm}^{\mu}(\rho)=\gamma_{\mu}\left[(n_{\mu}\left(\omega_{jm}\right)+1)\rho_{mm}-n_{\mu}\left(\omega_{jl}\right)\rho_{jj}\right]
\left|\left\langle \epsilon_{k}\right|S_{\mu}\left|\epsilon_{m}\right\rangle \right|^{2}
\end{equation}
with $\omega_{jm}=\epsilon_{m}-\epsilon_{j}$ representing the increment rate of the population
$\rho_{jj}$. The evolution of the density matrix $(\ref{differential equation})$ can rewrite as
 $\frac{d \vert \rho_{ss} \rangle}{dt}=M \vert \rho_{ss} \rangle$ with 
$\vert \rho_{ss} \rangle=[\rho_{11}, \rho_{22}, \rho_{33},\rho_{44}]^{T}$ and the matrix $M=\sum_{\mu} M_{\mu}$ \cite{PhysRevE.90.042142}. The matrix $M_{\mu}$ is expressed as 
\begin{align}
M_{\mu}=\left(\begin{array}{cccc}
M^{\mu}_{11} &  A_{\mu 2} &  A_{L4} & A_{\mu 6} \\
B_{\mu 2} & M^{\mu}_{22} & B_{\mu 1} &  B_{\mu 3} \\
B_{\mu 4} & A_{\mu 1} & M^{\mu}_{33} & B_{\mu 5} \\
B_{\mu 6} &A_{\mu 3} &  A_{\mu 5} & M^{\mu}_{44}
\end{array}\right),
\end{align}
where $M^{\mu}_{11}=-\sum_{l=2, 4, 6} B_{\mu l}$, $M^{\mu}_{22}=-\sum_{l=1}^{3} A_{\mu l}$, $M^{\mu}_{33}=-  B_{\mu 1} -\sum_{l=4, 5} A_{\mu l}$, and $M^{\mu}_{44}=-\sum_{l=3, 5} B_{\mu 3}- A_{\mu 6}$ with $A_{\mu l}=2 J_\mu\left(-\omega_{\mu l}\right) a_{\mu l}^{2}$, $B_{\mu l}=2 J_\mu\left(\omega_{\mu l}\right) a_{\mu l}^{2}$. 
Let $\frac{d \vert \rho_{ss} \rangle}{dt}=0$, one will obtain the steady-state $\rho_{ss}$.  
Here, we focus on the thermodynamical behavior in the long time limit; the definition of heat current can be considered as follows \cite{breuer2002theory}
\begin{align} \label{heatcurrent}
\dot{\mathcal{Q}}_{\mu}=\mathrm{Tr}\lbrace H_S \mathcal{L}[\rho_{ss}]\rbrace,
\end{align}
where $\dot{\mathcal{Q}}_{\mu}>0$ denotes heat current transfer from reservoir $\mu$ to system. The expression of heat current Eq. (\ref{heatcurrent}) can also expressed as 
\begin{align} \label{Heat current}
\dot{\mathcal{Q}}_{\mu}=-\sum_{j m} \Gamma^{\mu}_{j m}(\rho_{ss}) \omega_{j m}=\langle \epsilon \vert M_{\mu} \vert \rho_{ss}\rangle,
\end{align}
where transition energy $\omega_{j m}=\epsilon_{m}-\epsilon_{j}$, and the transition rates $\Gamma^{\mu}_{j m}(\rho_{ss})= A_{\mu l} \rho_{jj} -B_{\mu l} \rho_{mm}$ defined before. 

 \begin{figure}[!htbp]
\centering \includegraphics[width=1\columnwidth]{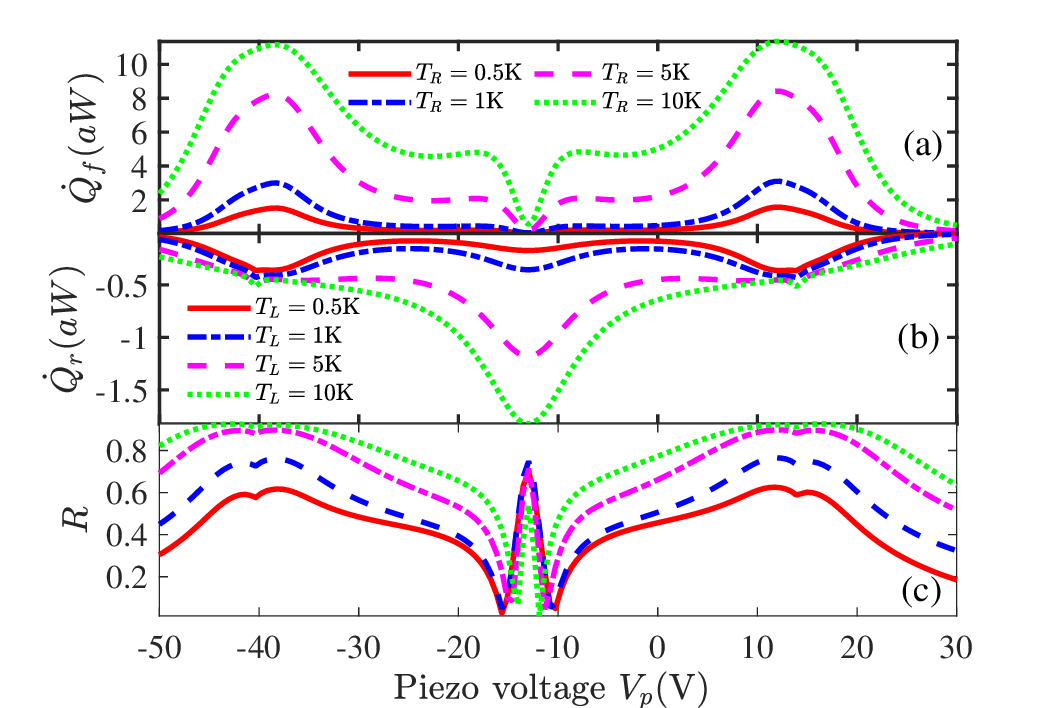
} \caption{The forward (a), reverse (b) heat currents, and rectification coefficient (c) versus the applied piezo voltage. The related parameters: dissipation rates $\gamma_{L}/2\pi=3 \mathrm{MHz}$, $\gamma_{R}=\gamma_{L}$ and temperatures $T_{L/R}=0.1 \mathrm{K}$ for the forward and reverse transfer, respectively, and other parameters are the same as in Fig. \ref{Fig_frequency}. }
\label{heatcurrent_V_p1} 
\end{figure}
As for a quantum heat diode, its performance can be  characterized by a rectification factor, and it is defined as \cite{PhysRevE.90.042142}
\begin{align}
R=\frac{|\dot{\mathcal{Q}}_{f}+\dot{\mathcal{Q}}_{r}|}{|\dot{\mathcal{Q}}_{f}-\dot{\mathcal{Q}}_{r}|}.
\end{align}
We refer to the forward (reverse) heat current as $\dot{\mathcal{Q}}_{f/r}=\dot{\mathcal{Q}}_{R}$ and  $\dot{\mathcal{Q}}_{f}\geq 0$, if $T_{R}\geq T_{L}$, and vice versa. This factor reflects the imbalance between the
two directions of heat current when exchanging temperatures of two terminals. 
The range of $R$ can take $0\leq R \leq 1$; there is no rectification if $R=0$, a perfect diode when $R=1$, and a well-performance diode when $0<R<1$.

\section{Results and discussions}
\label{Results and discussions}
 In the simulation, we consider a strong internal coupling regime $g \sim 850\mathrm{MHz}$ and follow the experimental parameters of Refs.  \cite{lisenfeld2015observation, de2022strain, lisenfeld2016decoherence, PhysRevB.95.241409} unless we stress. The related system parameters are: the tunneling energy $\Delta_L=7.5 \mathrm{GHz}$, $\Delta_R=1.3 \mathrm{GHz}$; the asymmetry energy $\varepsilon_L=c_L V_p-3.3[\mathrm{GHz}]$, $\varepsilon_R\left(V_p\right)=c_R\left(V_p+13[\mathrm{V}]\right)$, $c_L=5 \mathrm{MHz} \mathrm{V}^{-1}$ $c_R=0.3 \mathrm{GHz} \mathrm{V}^{-1}$; the internal coupling strength $g=850 \mathrm{MHz}$. As shown in Fig. \ref{Fig_frequency}, we plot the frequencies $\omega_{\mu}$ versus the Piezo voltage $V_p$. It is found that $\omega_{L}$ is nearly kept constant, but $\omega_{R}$ increases first and then decreases as Piezo voltage $V_p$  increases. Under such a parameter condition,
 modulating $V_p$ can significantly change the frequency $\omega_R$. Thus, one can modulate the voltage to change the frequency and further modulate the heat currents. We plot the heat current as a function of Piezo voltage shown in Fig. \ref{heatcurrent_V_p1} (a) and Fig. \ref{heatcurrent_V_p1} (b). 
From Fig. \ref{heatcurrent_V_p1} (a) and Fig. \ref{heatcurrent_V_p1} (b), we find that heat currents show non-monotonic behavior, and the variation of temperature bias can enhance the magnitude of heat transfer, i.e., the higher temperature bias, the larger heat current. 
The maximal heat transport in Fig. \ref{heatcurrent_V_p1} (a) can be realized at $V_p=-40.3 \mathrm{V}$ and $V_p=13.9 \mathrm{V}$ for resonant case $\omega_{L}=\omega_{R}$. It indicates the resonant condition not only realizing the large energy exchange with thermal reservoirs but also achieving the large rectification effects, which is similar to Ref. \cite{PhysRevE.99.042121}. When we consider the reverse heat transport, heat currents closely resembled the forward case in low temperatures with $T_{L}=0.5 \mathrm{K}, 1\mathrm{K}$. Still, the higher temperatures at $T_{L}=5 \mathrm{K}$, $10 \mathrm{K}$ will change the position of heat current maximums from $V_p=-40.3 \mathrm{V}$ and $V_p=13.9 \mathrm{V}$ to $V_p=-13 \mathrm{V}$. Hence, one can modulate the Piezo voltage $V_p$ to control the magnitude of heat currents, and this system can realize the function of a heat valve. What is more important, it is also found that heat currents exhibit nonreciprocal heat transfer from Fig. \ref{heatcurrent_V_p1} (a) and (b), which means that this system can also be implemented as a thermal diode. The rectification factor can characterize the performance of the nonreciprocal heat transport rectification factor. The rectification factor versus voltage $V_p$ is shown in Fig. \ref{heatcurrent_V_p1} (c), and nearly perfect rectification can be approximately obtained at $V_p \approx -40.3 \mathrm{V}$ and $V_p \approx 13.9\mathrm{V} $ at large temperature difference. However, at the two points $V_p\approx -14V,-11V$,  the rectification effects approach zero.  This can be understood as follows.  Note that $\gamma_R=\gamma_L$ isn't enough to mean the symmetric system-reservoir coupling because of the potential different transition channels and operators as shown in Eq. ({\ref{the eigenoperators}}). In this sense, one can roughly understand the asymmetry of atomic interaction and the asymmetry of system-reservoir couplings could offset, which leads to the reciprocal heat transfer at the two particular points. A subtle analysis is given in Appendix \ref{CC}. 

 \begin{figure}[!htbp]
\centering \includegraphics[width=1\columnwidth]{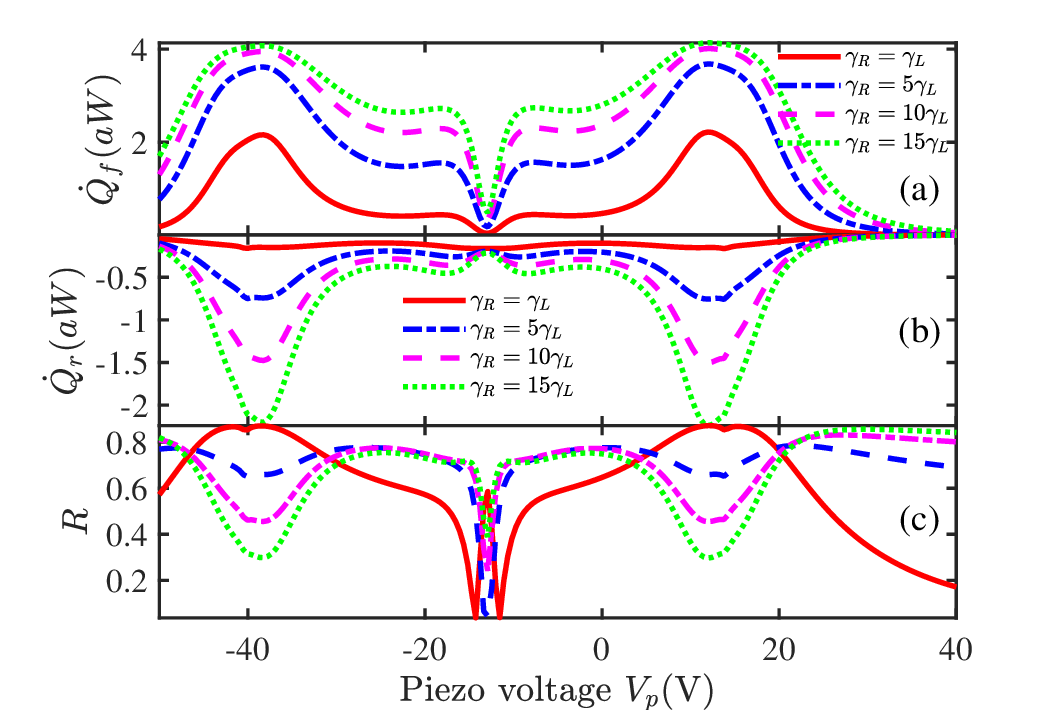} \caption{The forward (a), reverse (b) heat currents, and rectification coefficient (c) versus the applied voltage for different system-reservoir couplings. System parameters: dissipation rates $\gamma_{L}/2\pi=1 \mathrm{MHz}$; In forward process, temperatures take $T_{L}=0.1 \mathrm{K}$, $T_{R}=3 \mathrm{K}$; In reverse process $T_{L}=0.1 \mathrm{K}$, $T_{R}=3 \mathrm{K}$, and other parameters are the same as in Fig. \ref{Fig_frequency}.}
\label{heatcurrent_V_p2} 
\end{figure}

\begin{figure}[!htbp]
\centering
\subfigure[]{
 \includegraphics[width=0.48\columnwidth]{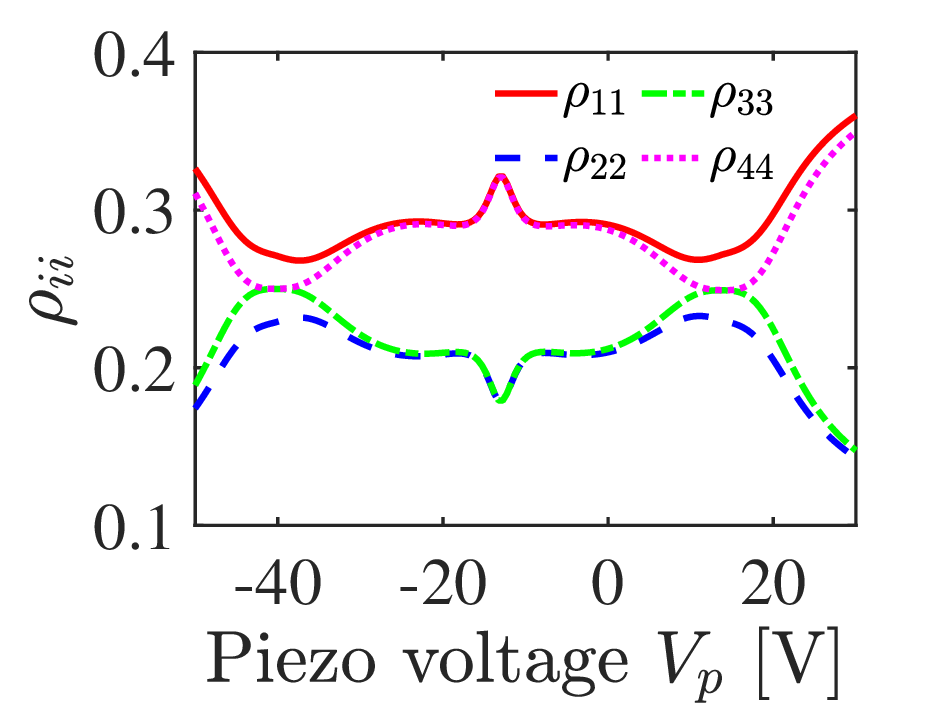
} }\hspace{-0.5cm}
\subfigure[]{
 \includegraphics[width=0.48\columnwidth]{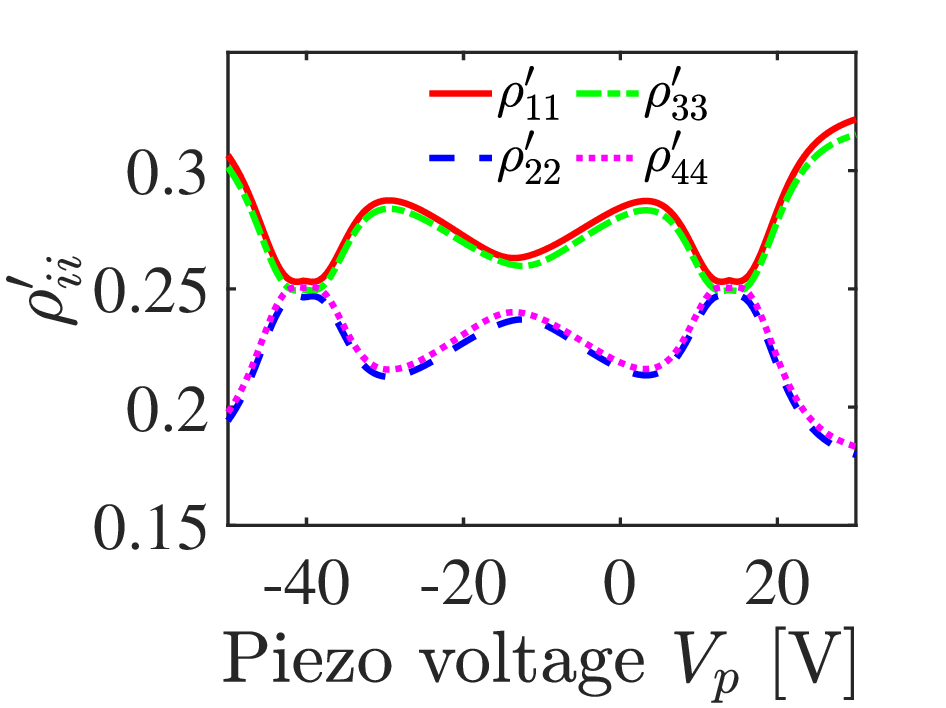
} }
\caption{The forward (a) and reverse (b) populations as a function of the Piezo voltage when considering the different coupling strengths of system-reservoir. Parameters: dissipation rates $\gamma_{L}/2\pi=1 \mathrm{MHz}$, $\gamma_{R}=5 \gamma_{L}$; $T_{L}=0.1 \mathrm{K}$, $T_{R}=5 \mathrm{K}$ for forward process; $T_{R}=5 \mathrm{K}$, $T_{L}=0.1 \mathrm{K}$ for reverse process and other parameters as shown in Fig. \ref{Fig_frequency}.}
\label{Population} 
\end{figure}

\begin{figure}[!htbp]
\centering \includegraphics[width=1\columnwidth]{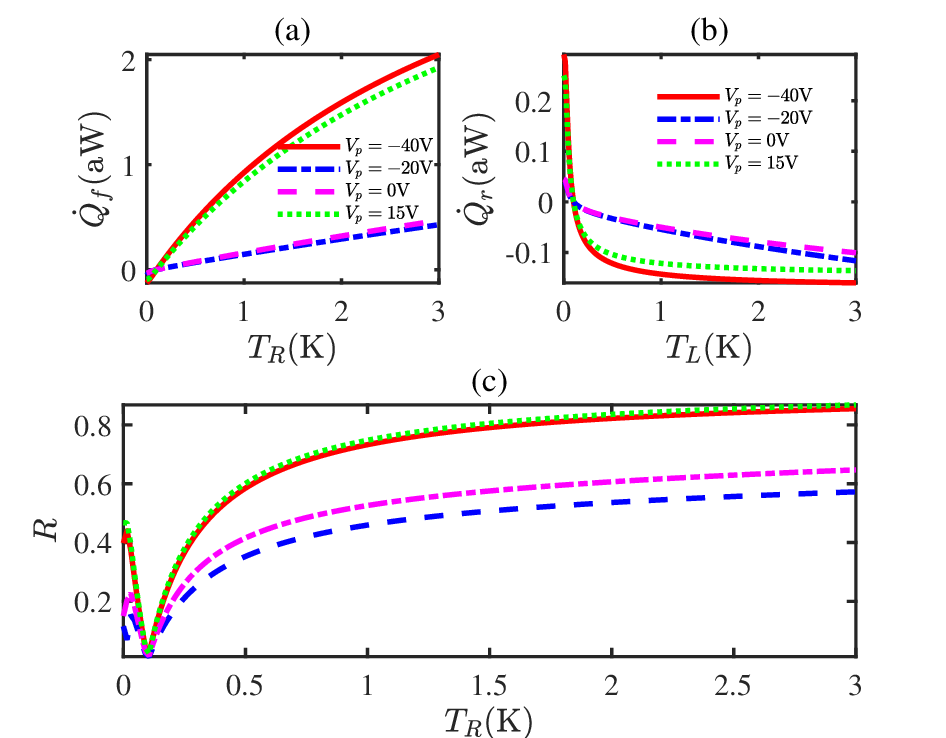
} \caption{The forward (a), reverse (b) heat currents, and rectification coefficient (c) versus the temperatures for different voltage $V_p$. System parameters: dissipation rates $\gamma_{L}/2\pi=3 \mathrm{MHz}$, $\gamma_{R}=\gamma_{L}$; In forward process, temperatures take $T_{L}=0.1 \mathrm{K}$; In reverse process $T_{R}=0.1 \mathrm{K}$, and other parameters are the same as in Fig. \ref{Fig_frequency}.}
\label{heatcurrent_tem} 
\end{figure}

As mentioned previously, different system-reservoir couplings are an important factor in realizing a heat rectification effect \cite{PhysRevLett.94.034301, senior2020heat}. Next, we will show whether the performance of the thermal diode can be boosted using the different system-reservoir coupling strengths in Fig. \ref{heatcurrent_V_p2}. It is found that the stronger asymmetric coupling strength can realize the larger heat transport. Moreover, the performance of heat diodes employing asymmetric and symmetric couplings can be complementary. In comparison of the equal system-reservoir coupling strength, for $\omega_{R}=\omega_{L}$ the rectification effects at when $V_p=-40.3 \mathrm{V}$ and $V_p=13.9 \mathrm{V}$ are weakened, which can also be attributed to the competition of the asymmetry of atomic interaction and the asymmetry of the system-reservoir couplings. The rectification at $V_p\approx -13 \mathrm{V}$ in Fig. \ref{heatcurrent_V_p2} can also be understood similarly.   In addition, the nearly stable rectification effect can be obtained at $V_p \in [-30, -15], [-10,5], [20,40] \mathrm{V}$. To explain this result, we give the population number as a function of $V_p$ as shown in Fig. \ref{Population}. We find that in these ranges, populations $\rho_{11}$, $\rho_{44}$ are dominantly for forward heat transfer, while $\rho^{\prime}_{11}$, $\rho^{\prime}_{33}$ play the main role for reverse process. It means that the transition channel $\left|\epsilon_1\right\rangle \leftrightarrow\left|\epsilon_4\right\rangle$, $\left|\epsilon_2\right\rangle \leftrightarrow\left|\epsilon_1\right\rangle$ of forward transfer is different reverse process $\left|\epsilon_1\right\rangle \leftrightarrow\left|\epsilon_3\right\rangle$, $\left|\epsilon_2\right\rangle \leftrightarrow\left|\epsilon_1\right\rangle$, and it may enable directional heat flow. Besides we also researched the temperature-modulated heat valve and diode with different voltages $V_p$. From Fig. \ref{heatcurrent_tem}, we find that there is no heat current when temperatures of two terminals are equal $T_{R}=T_{L}$, and the higher temperature bias is beneficial to heat rectification. The rectification effect's mechanism stems from the asymmetry of energy structure, as shown in Fig. \ref{Fig_frequency}. In summary, one can realize a well-performance diode in a wide parameter range.

 \begin{figure}[!htbp]
\centering
\subfigure[]{
 \includegraphics[width=0.48\columnwidth]{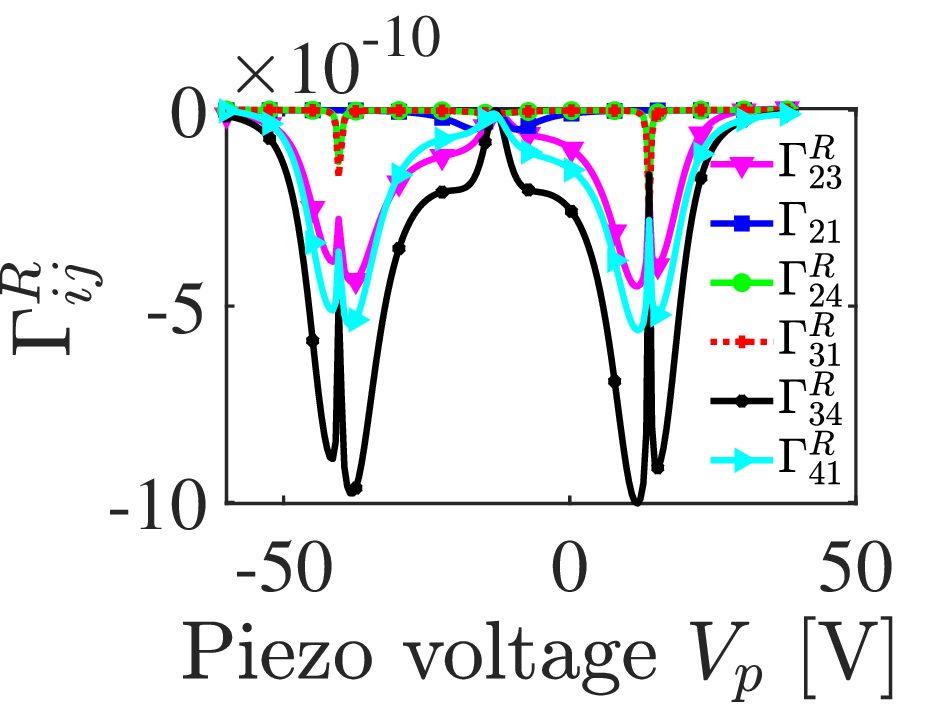
} } \hspace{-0.5cm}
\subfigure[]{
 \includegraphics[width=0.48\columnwidth]{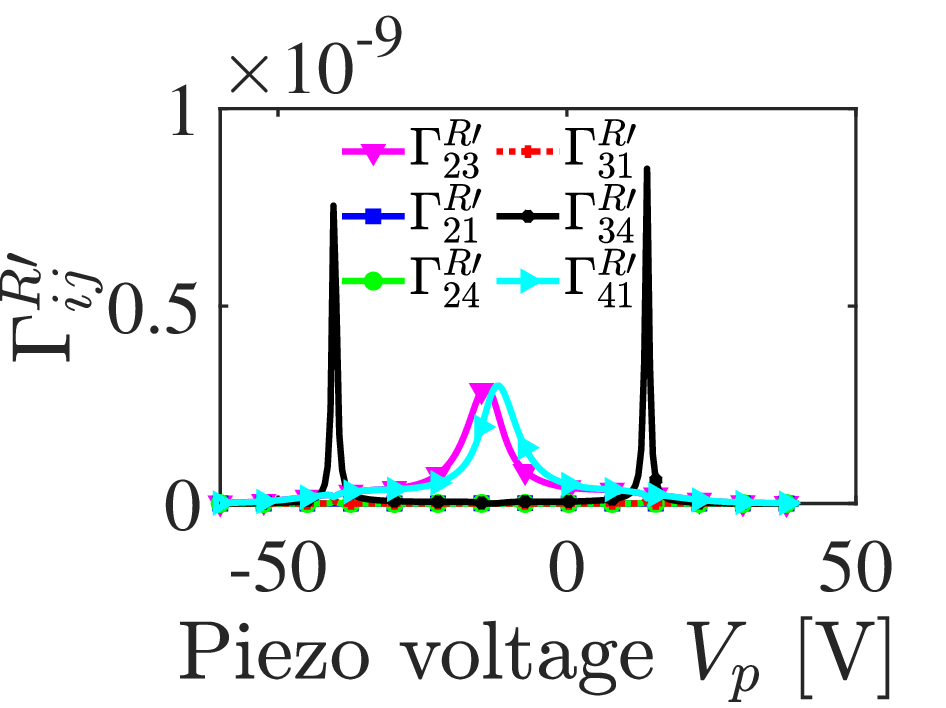
} }
\caption{The transition rates for the forward (a) and reverse (b) versus the applied Piezo voltage. Parameters: $\gamma_{L}/2\pi=3 \mathrm{MHz}$, $\gamma_{R}=\gamma_{L}$; $T_{L}=0.1 \mathrm{K}$, $T_{R}=5 \mathrm{K}$ for forward process; $T_{R}=5 \mathrm{K}$, $T_{L}=0.1 \mathrm{K}$ for reverse process and other parameters as shown in Fig. \ref{Fig_frequency}.}
\label{Flux_two_qubits_transition} 
\end{figure}

 \begin{figure}[!htbp]
\centering
\subfigure[]{
 \includegraphics[width=0.48\columnwidth]{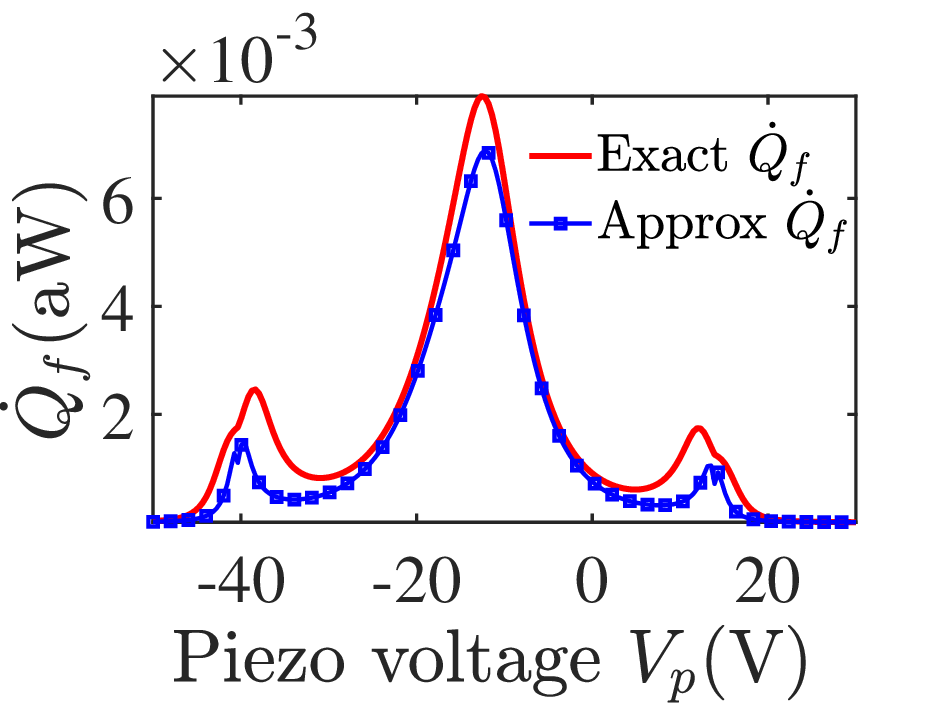
} } \hspace{-0.5cm}
\subfigure[]{
 \includegraphics[width=0.48\columnwidth]{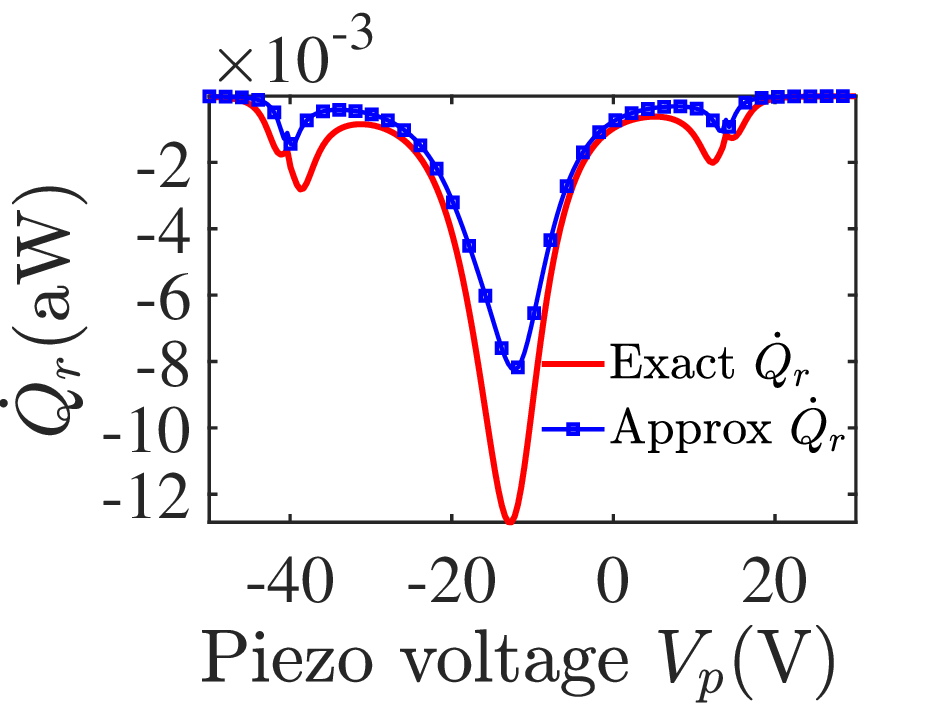
} }
\caption{Heat currents as a function of Piezo voltage $V_p$ for forward (a) and reverse (b) heat transfer. The red solid line and blue squared line denote the exact (cf. Eq. (\ref{Heat current})), approximate heat currents (cf. Eq. (\ref{appro-heat current})), respectively. Here $\gamma_{L}/2\pi=3 \mathrm{MHz}$, $\gamma_{R}=\gamma_{L}$; temperatures take $T_L=5 \mathrm{m K}$, $T_R=10 \mathrm{m K}$ for forward transfer, otherwise, exchanging temperature of two terminals of the reservoir. Other parameters are the same as shown in Fig. \ref{Fig_frequency}.}
\label{Flux_two_qubits_heat} 
\end{figure}

The occurrence of the rectification effect requires some asymmetry of the total system. No rectification effect exists when the total system, including heat reservoirs, is fully symmetric, i.e.,  the same asymmetry energy $\varepsilon_{\mu}$, tunneling energy $\Delta_{\mu}$ and system-reservoir couplings for the current system.  To explain the reason for the heat rectification effect shown in Fig. \ref{heatcurrent_tem}. The existence of different energy structures of atoms induced by modulation of Piezo voltage and different tunneling provides the asymmetry to realize nonreciprocal heat transport. As shown in Fig. \ref{Fig_frequency} (b). It involves the same six transition channels for each atom with independent reservoir $\left|\epsilon_2\right\rangle \leftrightarrow\left|\epsilon_3\right\rangle$, $\left|\epsilon_2\right\rangle \leftrightarrow\left|\epsilon_1\right\rangle$, $\left|\epsilon_2\right\rangle \leftrightarrow\left|\epsilon_4\right\rangle$, $\left|\epsilon_3\right\rangle \leftrightarrow\left|\epsilon_1\right\rangle$, $\left|\epsilon_3\right\rangle \leftrightarrow\left|\epsilon_4\right\rangle$, and $\left|\epsilon_4\right\rangle \leftrightarrow\left|\epsilon_1\right\rangle$, but the coefficient of transition rates are different (cf. Eq.(\ref{the eigenoperators})). Meanwhile, the two terminals have different temperatures, providing the various transition abilities of the same channel. For example, we take a transition channel $\left|\epsilon_1\right\rangle \leftrightarrow\left|\epsilon_4\right\rangle \leftrightarrow\left|\epsilon_3\right\rangle \leftrightarrow\left|\epsilon_2\right\rangle \leftrightarrow\left|\epsilon_1\right\rangle$. When we consider the forward transfer $T_{L}<T_{R}$, the higher temperature $T_R$ can realize $\left|\epsilon_1\right\rangle \rightarrow\left|\epsilon_4\right\rangle \rightarrow\left|\epsilon_3\right\rangle \rightarrow\left|\epsilon_2\right\rangle$, however, when we invert the temperatures of the two terminals, the lower temperature $T_{R}$ have insufficient ability to fulfill it. From this perspective, it is found that the reverse heat transport is a blockade. From Eq. (\ref{Heat current}), heat current is the sum of the product of transition rate and transition energy. To further understand the rectification effect, we plot the transition rates as a function of piezo voltage shown in Fig. \ref{Flux_two_qubits_transition}. We are comparing Fig. \ref{Flux_two_qubits_transition} (a) and (b); the transition rates with the same transition energy are different for forward and reverse transfer, which causes nonreciprocal heat transport. Here, we also consider it analytically and obtain a solution at low temperatures shown in Appendix \ref{Appendix A}. Fig. \ref{Flux_two_qubits_heat} exhibits approximate heat current (cf. Eqs. (\ref{appro-heat current}) for forward and reverse heat transfer in comparison to the exact case (cf. Eq. (\ref{Heat current})). Although the magnitude of the heat current is small, it provides an analytical form to understand the nonreciprocal heat transport. The approximate heat currents are valid in a wide region from this figure. A heat valve can realize this regardless of the forward and reverse transfer. One also finds that nonreciprocal heat transport can occur when we invert the temperatures of two terminals. From Eq. (\ref{appro-heat current}), one can find that it provides the same result when exchanging the temperatures of two terminals.

Finally, we would like to consider the proposal's feasibility. The two coupling two-level system defects can realized in in the AlOx of Josephson junction tunnel barriers. Each atom can coupled to a normal-metal resistor \cite{senior2020heat} or general noiseless resistor with a
fluctuating voltage source \cite{RevModPhys.82.1155, https://doi.org/10.1002/qute.202100054} act as a thermal bath. The setup of tuning a two-level system by applied mechanical strain has been extensively discussed in Refs. \cite{bilmes2021quantum, grabovskij2012strain}. The two-level system relaxation rates $\gamma_{\mu}/2 \pi$ can be taken in the range $0.1$-$5 \mathrm{MHz}$ \cite{PhysRevLett.110.193602}. In addition, heat current can be detected by employing the normal-metal-insulator-superconductor(NIS) thermometry techniques \cite{ronzani2018tunable, RevModPhys.78.217}.

\section{Conclusion and discussion}
\label{conclusion}
We have theoretically proposed a heat valve and diode via modulating the Piezo voltage in a defect coupling atoms system. It is shown that a well-performance heat valve and diode can be realized at an extensive parameter range. We discuss the effect of Piezo voltage, temperature, and system-reservoir coupling strength on the performance of heat devices and give a physical explanation for the phenomena. We hope this work can provide an efficient way to design and modulate quantum thermal rectification devices.

\section*{Acknowledgements}
This work was supported by the National Natural Science Foundation of China under Grants No.12175029, No.
12011530014 and No.11775040.

\appendix

\section{The eigenoperator of open system} \label{Appendix A}

 In $H_S$ representation, the jump operators $S_{\mu}$ of system can be solved as
\begin{align}  \label{the eigenoperators}
\nonumber
S_{L 1} &=\sin \theta_{L} \sin(\frac{\alpha+\beta}{2}) \left\vert \epsilon_{3} \right\rangle \left\langle \epsilon_{2} \right\vert,  \\ \nonumber
S_{L 2} &=-\cos \theta_{L} \sin \alpha\left\vert\epsilon_{1}\right\rangle\left\langle\epsilon_{2}\right\vert, \\\nonumber
S_{L 3} &=\sin \theta_{L} \cos(\frac{\alpha+\beta}{2})\left\vert\epsilon_{4}\right\rangle\left\langle\epsilon_{2}\right\vert, \\\nonumber
S_{L 4} &=\sin \theta_{L} \cos(\frac{\alpha+\beta}{2})\left\vert\epsilon_{1}\right\rangle\left\langle\epsilon_{3}\right\vert, \\\nonumber
S_{L 5} &=-\cos \theta_{L} \sin \beta \left\vert\epsilon_{4}\right\rangle\left\langle\epsilon_{3}\right\vert, \\\nonumber
S_{L 6} &=-\sin \theta_{L} \sin(\frac{\alpha+\beta}{2}) \left\vert\epsilon_{1}\right\rangle\left\langle\epsilon_{4}\right\vert,  \\
S_{R 1} &=\sin \theta_{R} \cos(\frac{\alpha-\beta}{2}) \left\vert\epsilon_{3}\right\rangle\left\langle\epsilon_{2}\right\vert, \\\nonumber
S_{R 2} &=-\cos \theta_{R} \sin \alpha \left\vert\epsilon_{1}\right\rangle\left\langle\epsilon_{2}\right\vert, \\\nonumber
S_{R 3} &=\sin \theta_{R} \sin(\frac{\alpha-\beta}{2})\left\vert\epsilon_{4}\right\rangle\left\langle\epsilon_{2}\right\vert, \\\nonumber
S_{R 4} &=-\sin \theta_{R} \sin(\frac{\alpha-\beta}{2})\left\vert\epsilon_{1}\right\rangle\left\langle\epsilon_{3}\right\vert, \\\nonumber
S_{R 5} &=\cos \theta_{R} \sin \beta \left\vert\epsilon_{4}\right\rangle\left\langle\epsilon_{3}\right\vert, \\\nonumber
S_{R 6} &=\sin \theta_{R} \cos(\frac{\alpha-\beta}{2}) \left\vert\epsilon_{1}\right\rangle\left\langle\epsilon_{4}\right\vert, \\\nonumber
S_{L 0} &=\cos \theta_{L} \cos \alpha(\left\vert\epsilon_{3}\right\rangle\left\langle\epsilon_{3}\right\vert-\left\vert\epsilon_{2}\right\rangle\left\langle\epsilon_{2}\right\vert)\\\nonumber&+\cos \theta_{L} \cos \beta (\left\vert\epsilon_{1}\right\rangle\left\langle\epsilon_{1}\right\vert-\left\vert\epsilon_{4}\right\rangle\left\langle\epsilon_{4}\right\vert), \\\nonumber
S_{R 0} &=\cos \theta_{R} \cos \alpha (\left\vert\epsilon_{3}\right\rangle\left\langle\epsilon_{3}\right\vert -\left\vert\epsilon_{2}\right\rangle\left\langle\epsilon_{2}\right\vert) \\\nonumber&+\cos \theta_{R} \cos \beta (\left\vert\epsilon_{4}\right\rangle\left\langle\epsilon_{4}\right\vert-\left\vert\epsilon_{1}\right\rangle\left\langle\epsilon_{1}\right\vert). 
\end{align}
The corresponding eigenfrequencies are 
$\omega_{\mu 1}=\omega_{32}=\epsilon_{2}-\epsilon_{3}$, 
$\omega_{\mu 2}=\omega_{12}=\epsilon_{2}-\epsilon_{1}$,
$\omega_{\mu 3}=\omega_{42}=\epsilon_{2}-\epsilon_{4}$,
$\omega_{\mu 4}=\omega_{13}=\epsilon_{3}-\epsilon_{1}$,
$\omega_{\mu 5}=\omega_{43}=\epsilon_{3}-\epsilon_{4}$,
$\omega_{\mu 6}=\omega_{14}=\epsilon_{4}-\epsilon_{1}$.
Here we tab the coefficient of $S_{\mu l}$ as $a_{\mu l}$, for example, $a_{L1}=\sin \theta_{L} \sin(\frac{\alpha+\beta}{2})$.

\section{The population, and heat current for coupled system } \label{Appendix B}
\begin{widetext} 
The density matrix for the composite system at steady-state can be solved as 
\begin{align} \label{rhoss}
&\rho_{11}=\nonumber ((A_4+A_5+B_1)(A_3 A_6+A_2 (A_6+B_3))+(A_3 A_4+A_2 (A_4+B_1)) B_5+A_1 (A_5 A_6+A_4(A_6+B_3+B_5)))/N, \\ \nonumber
&\rho_{22}=\nonumber (A_5 A_6 B_2+(A_4 B_2+B_1 (B_2+B_4))(A_6+B_3+B_5)+A_4 B_3 B_6+B_1 (B_3+B_5) B_6+A_5 B_3 (B_2+ B_4+B_6))/N, \\ \nonumber
&\rho_{33}=\nonumber (A_3 A_6 B_4+(A_2 B_4+A_1 (B_2+B_4))(A_6+B_3+B_5)+A_2 B_5 B_6+A_1 (B_3+B_5) B_6+A_3 B_5(B_2+B_4+B_6))/N,\\ \nonumber
&\rho_{44}=\nonumber (A_2 A_5 B_4+A_1 A_5 (B_2+B_4)+(A_1+A_2)(A_4+A_5) B_6+A_2 B_1 B_6+A_3 A_4 (B_2+B_6)+A_3(A_5+B_1)(B_2+B_4+\\ &B_6))/N.
\end{align}
Here $N$ denotes the normalized coefficient, $A_{i}=\sum_{\mu} A_{\mu i}$ and $B_{i}=\sum_{\mu} B_{\mu i}$.
To obtain an analytical expression, we consider low temperatures. It means $e^{-\frac{\omega}{T_{L}}}, e^{-\frac{\omega}{T_{R}}} \ll 1$ and the involved transition are $\left|\epsilon_2\right\rangle \leftrightarrow\left|\epsilon_3\right\rangle$, $\left|\epsilon_3\right\rangle \leftrightarrow\left|\epsilon_4\right\rangle$, and $\left|\epsilon_4\right\rangle \leftrightarrow\left|\epsilon_1\right\rangle$. For simplification, we take $\gamma=\gamma_{L}$, and $\kappa=\gamma_{R}$. Based on those conditions, one can simplify Matrix $M$ with normalized condition $\mathrm{Tr}(\rho)=1$ as 
\begin{align}
M_{m}=\left(\begin{array}{cccc}
M_{m}^{11} & 0 & 0 & 2 a_{L6}^2 \gamma+2 a_{R6}^2 \kappa \\
0 & -2 a_{L1}^2  \gamma-2 a_{R1}^2 \kappa & 2 a_{L1}^2 e^{\frac{-\omega_{32}}{T_L}} \gamma+2 a_{R1}^2 e^{\frac{-\omega_{32}}{T_R}} \kappa & 0 \\
2 a_{R4}^2 e^{\frac{-\omega_{13}}{T_R}} \kappa  &2 a_{L1}^2  \gamma+2 a_{R1}^2  \kappa &M^{33}_{m}&2 a_{L5}^2 e^{\frac{-\omega_{43}}{T_L}} \gamma+2 a_{R5}^2 e^{\frac{-\omega_{43}}{T_R}} \kappa  \\
1 & 1 & 1 & 1
\end{array}\right),
\end{align}
where $M_{m}^{11}=-2 a_{R4}^2 e^{\frac{-\omega_{13}}{T_R}} \kappa -2 a_{R6}^2 e^{-\frac{-\omega_{14}}{T_R}} \kappa$, and $M^{33}_{m}=-2 a_{L5}^2 \gamma-2 a_{R5}^2 \kappa -2 a_{L1}^2 e^{\frac{-\omega_{32}}{T_L}} \gamma-2 a_{R1}^2 e^{\frac{-\omega_{32}}{T_R}} \kappa$.
The matrix determinant of $M_{m}$ can be expressed as
\begin{align}
\begin{split} 
\mathrm{Det}(M_m) &=8 e^{-\frac{(T_L+T_R) (\epsilon _2+\epsilon _3+\epsilon _4)}{T_L T_R}} (-(e^{\frac{\epsilon _2}{T_R}+\frac{\epsilon
_3}{T_L}} \gamma a_{L1}^2+e^{\frac{\epsilon _2}{T_L}+\frac{\epsilon_3}{T_R}} \kappa a_{R1}^2) ((e^{\frac{\epsilon
_3}{T_R}+\frac{\epsilon _4}{T_L}} \gamma a_{L5}^2+e^{\frac{\epsilon_3}{T_L}+\frac{\epsilon_4}{T_R}} \kappa a_{R5}^2)(e^{\frac{\epsilon_1}{T_L}+\frac{\epsilon _4}{T_R}} \gamma a_{L6}^2+\\  &  
(e^{\frac{\epsilon_1}{T_L}+\frac{\epsilon _4}{T_R}} \gamma a_{L6}^2+ 
e^{\frac{\epsilon_1}{T_R}+\frac{\epsilon_4}{T_L}}(e^{\frac{\epsilon_1}{T_L}+\frac{\epsilon _4}{T_R}} \gamma a_{L6}^2+(e^{\frac{\epsilon_1}{T_L}+\frac{\epsilon _4}{T_R}} \gamma a_{L6}^2+e^{\frac{\epsilon _1}{T_R}+\frac{\epsilon_4}{T_L}}\kappa a_{\text{R6}}^2)-e^{\frac{(T_L+T_R) \epsilon _3}{T_L T_R}} (\gamma a_{L1}^2+\kappa  a_{R1}^2)\\  &
(e^{\frac{\epsilon_4}{T_R}} (e^{\frac{\epsilon_1}{T_L}}+e^{\frac{\epsilon _4}{T_L}}) \gamma a_{L6}^2+e^{\frac{\epsilon
_4}{T_L}} (e^{\frac{\epsilon _1}{T_R}}+e^{\frac{\epsilon_4}{T_R}}) \kappa  a_{R6}^2))-(\gamma a_{L1}^2+\kappa  a_{R1}^2) (e^{\frac{(T_L+T_R) \epsilon _2}{T_L T_R}} (e^{\frac{\epsilon
_3}{T_R}+\frac{\epsilon _4}{T_L}}\gamma  a_{L5}^2+e^{\frac{\epsilon _3}{T_L}+\frac{\epsilon _4}{T_R}} \kappa  a_{R5}^2) \\ &
(e^{\frac{\epsilon_1}{T_L}+\frac{\epsilon_4}{T_R}} \gamma  a_{L6}^2+e^{\frac{\epsilon _1}{T_R}+\frac{\epsilon_4}{T_L}}
\kappa  a_{\text{R6}}^2)+e^{\frac{(T_L+T_R) \epsilon_3}{T_L T_R}} (e^{\frac{\epsilon_2}{T_R}+\frac{\epsilon_3}{T_L}}\gamma a_{L1}^2+e^{\frac{\epsilon_2}{T_L}} (e^{\frac{\epsilon _3}{T_R}} \kappa  a_{R1}^2+e^{\frac{\epsilon_2}{T_R}}(\gamma  a_{L5}^2+\kappa  a_{R5}^2))) (e^{\frac{\epsilon_4}{T_R}} \\ &
 (e^{\frac{\epsilon_1}{T_L}}+e^{\frac{\epsilon
_4}{T_L}}) \gamma a_{L6}^2+e^{\frac{\epsilon_4}{T_L}} (e^{\frac{\epsilon _1}{T_R}}+e^{\frac{\epsilon_4}{T_R}})
\kappa a_{R6}^2))).
\end{split}
\end{align}

The corresponding steady state can be approximately solved as
\begin{align}
\begin{split}
\rho_{11}&=-\frac{8 (\gamma  a_{L1}^2+\kappa  a_{R1}^2) (\gamma a_{L5}^2+\kappa a_{R5}^2)
(\gamma a_{L6}^2+\kappa a_{R6}^2)}{\mathrm{Det}(M_m)},\\
\rho_{22}&=-\frac{8 e^{-\frac{(T_L+T_R) (\epsilon _2+\epsilon _3+\epsilon _4)}{T_L T_R}} (e^{\frac{\epsilon
_2}{T_R}+\frac{\epsilon _3}{T_L}} \gamma  a_{L1}^2+e^{\frac{\epsilon _2}{T_L}+\frac{\epsilon _3}{T_R}} \kappa  a_{R1}^2)
(e^{\frac{\epsilon _3}{T_R}+\frac{\epsilon _4}{T_L}} \gamma  a_{L5}^2+e^{\frac{\epsilon _3}{T_L}+\frac{\epsilon _4}{T_R}}
\kappa  a_{R5}^2) (e^{\frac{\epsilon _1}{T_L}+\frac{\epsilon _4}{T_R}} \gamma  a_{L6}^2+e^{\frac{\epsilon _1}{T_R}+\frac{\epsilon
_4}{T_L}} \kappa  a_{R6}^2)}{\mathrm{Det}(M_m)},\\
\rho_{33}&=-\frac{8 e^{-\frac{(T_L+T_R) (\epsilon _3+\epsilon _4)}{T_L T_R}} (\gamma  a_{L1}^2+\kappa
 a_{R1}^2) (e^{\frac{\epsilon _3}{T_R}+\frac{\epsilon _4}{T_L}} \gamma  a_{L5}^2+e^{\frac{\epsilon _3}{T_L}+\frac{\epsilon
_4}{T_R}} \kappa  a_{R5}^2) (e^{\frac{\epsilon _1}{T_L}+\frac{\epsilon _4}{T_R}} \gamma  a_{L6}^2+e^{\frac{\epsilon
_1}{T_R}+\frac{\epsilon _4}{T_L}} \kappa  a_{R6}^2)}{\mathrm{Det}(M_m)},\\
\rho_{44}&=-\frac{8 e^{-\frac{(T_L+T_R) \epsilon _4}{T_L T_R}} (\gamma  a_{L1}^2+\kappa  a_{R1}^2)
(\gamma  a_{L5}^2+\kappa  a_{R5}^2) (e^{\frac{\epsilon _1}{T_L}+\frac{\epsilon _4}{T_R}} \gamma  a_{L6}^2+e^{\frac{\epsilon
_1}{T_R}+\frac{\epsilon _4}{T_L}} \kappa  a_{R6}^2)}{\mathrm{Det}(M_m)}.
\end{split}
\end{align}
According to the definition of heat current Eq. (\ref{Heat current}) and the above equations, one can obtain an approximate expression as
\begin{align}  \label{appro-heat current}
\begin{split}
\dot{\mathcal{Q}}_{R}&=8 (2 e^{-\frac{(T_L+T_R) (\epsilon_2+\epsilon_3+\epsilon _4)}{T_L T_R}}  \kappa a_{R1}^2 (e^{\frac{\epsilon _2}{T_R}+
\frac{\epsilon _3}{T_L}}\gamma a_{L1}^2+e^{\frac{\epsilon_2}{T_L}+\frac{\epsilon
_3}{T_R}} \kappa  a_{R1}^2) (e^{\frac{\epsilon _3}{T_R}+\frac{\epsilon _4}{T_L}}  \gamma a_{L5}^2+e^{\frac{\epsilon
_3}{T_L}+\frac{\epsilon _4}{T_R}} \kappa a_{R5}^2)(e^{\frac{\epsilon_1}{T_L}+\frac{\epsilon_4}{T_R}} \gamma a_{L6}^2+\\& 
e^{\frac{\epsilon _1}{T_R}+\frac{\epsilon_4}{T_L}} \kappa a_{R6}^2) (\epsilon _2-\epsilon_3) +e^{-\frac{T_L \epsilon_2+(T_L+T_R) (\epsilon_3+\epsilon _4)}{T_L T_R}}\kappa  (\gamma  a_{L1}^2+\kappa a_{R1}^2)(e^{\frac{\epsilon _3}{T_R}+\frac{\epsilon _4}{T_L}}\gamma a_{L5}^2+e^{\frac{\epsilon _3}{T_L}+\frac{\epsilon_4}{T_R}} \kappa  a_{R5}^2)
(e^{\frac{\epsilon _1}{T_L}+\frac{\epsilon _4}{T_R}} \gamma  a_{L6}^2+\\ &
 e^{\frac{\epsilon
_1}{T_R}+\frac{\epsilon _4}{T_L}}\kappa  a_{R6}^2) (e^{\frac{\epsilon _3}{T_R}} a_{R1}^2 (-\epsilon
_2+\epsilon _3)+e^{\frac{\epsilon _2}{T_R}} a_{R5}^2 (\epsilon _3-\epsilon _4))+ 2 e^{\frac{\epsilon _1-\epsilon _4}{T_R}} \kappa  (\gamma  a_{L1}^2+\kappa a_{R1}^2) (\gamma  a_{L5}^2+\kappa
 a_{R5}^2)
 a_{R6}^2 (\gamma  a_{L6}^2+\\&\kappa  a_{R6}^2) (\epsilon _1-\epsilon _4)+2 e^{-\frac{T_L
\epsilon _3+(T_L+T_R) \epsilon_4}{T_L T_R}}\kappa (\gamma  a_{L1}^2+\kappa  a_{R1}^2) (\gamma a_{L5}^2+\kappa  a_{R5}^2) (e^{\frac{\epsilon _1}{T_L}+\frac{\epsilon _4}{T_R}}\gamma a_{L6}^2+e^{\frac{\epsilon
_1}{T_R}+\frac{\epsilon _4}{T_L}} \kappa a_{R6}^2) \\&
(-e^{\frac{\epsilon _3}{T_R}} a_{R6}^2 (\epsilon
_1-\epsilon _4)+e^{\frac{\epsilon _4}{T_R}} a_{R5}^2 (-\epsilon _3+\epsilon_4))/\mathrm{Det}(M_m).
\end{split}
\end{align}
\end{widetext}

\section{Heat rectification effects induced by the asymmetry of total system including environment. }\label{CC}
 \begin{figure}[!htbp]
\centering \includegraphics[width=1\columnwidth]{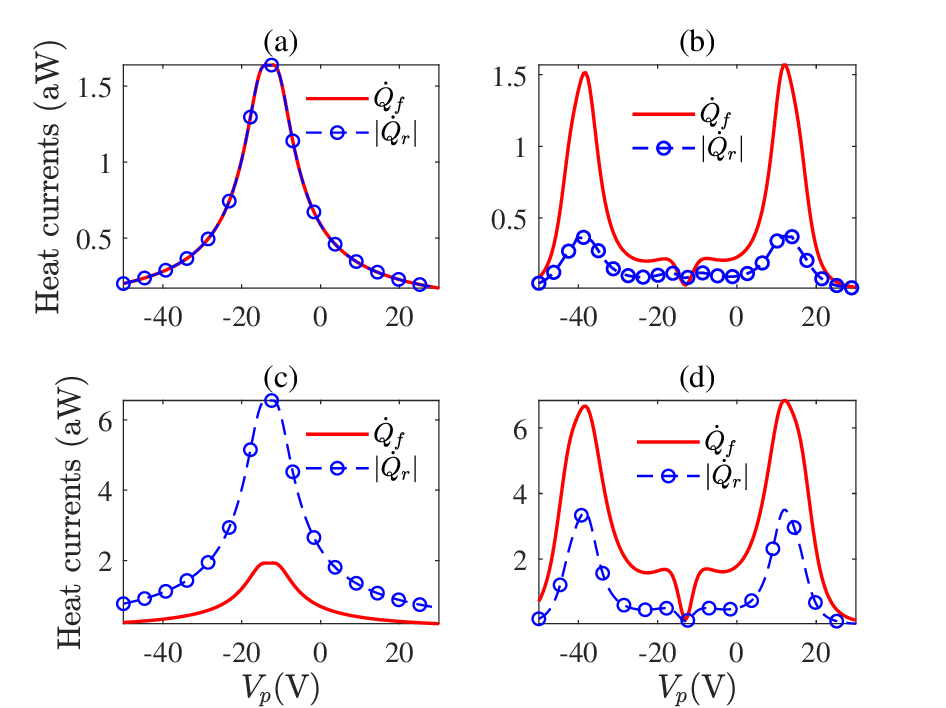
} \caption{The forward and reverse heat currents as a function of the applied piezo voltage $V_p$ with four different cases considering asymmetry of total system including environment. (a) Full symmetry, $\Delta_L=\Delta_R$, $\gamma_{R}=\gamma_{L}$, and $\varepsilon_R=\varepsilon_L$; (b) Different energy structures $\varepsilon_{\mu}$ and $\Delta_{\mu}$, $\varepsilon_L=c_L V_p-3.3[\mathrm{GHz}]$, $\Delta_L=7.5 \mathrm{GHz}$ ; (c) Different system-reservoir coupling strength $\gamma_{R}=10 \gamma_{L}$; (d) Different energy structures $\varepsilon_{\mu}$ and $\Delta_{\mu}$ and system-reservoir coupling strength $\gamma_{\mu}$, $\varepsilon_L=c_L V_p-3.3[\mathrm{GHz}]$, $\Delta_L=7.5 \mathrm{GHz}$, and $\gamma_{R}=10 \gamma_{L}$. The other parameters, In the forward process $T_{L}=0.1 \mathrm{K}$, $T_R=0.5 \mathrm{K}$; In the reverse process $T_{R}=0.1 \mathrm{K}$, $T_L=0.5 \mathrm{K}$; $g=850 \mathrm{MHz}$, $\Delta_R=1.3 \mathrm{GHz}$,  $\gamma_{L}/2\pi=3 \mathrm{MHz}$, $c_L=5 \mathrm{MHz} \mathrm{V}^{-1}$, $c_R=0.3 \mathrm{GHz} \mathrm{V}^{-1}$, and $\varepsilon_R\left(V_p\right)=c_R\left(V_p+13[\mathrm{V}]\right)$.  }
\label{Fig_Heat_diff} 
\end{figure}

We will analyze four cases for different total system structures to understand the thermal rectification effect more intuitively. In general, the asymmetry of the total system includes the asymmetry of subsystems and the asymmetry of the system-reservoir. As shown in Fig. \ref{Fig_Heat_diff}, we have plotted heat currents versus the $V_p$ with different asymmetry mechanisms. Compared to Fig. \ref{Fig_Heat_diff}(a), all other figures can have heat rectification effects. From Fig. \ref{Fig_Heat_diff}(b) and (d), we find that the different system energy structures in some ranges may lead to weak rectification, and different energy structures and system-reservoir coupling strength can produce competition in realizing heat rectification. Besides, one only considers the different system-reservoir coupling strengths shown in Fig. \ref{Fig_Heat_diff}(c), the rectification can also be enhanced.

\bibliography{main_text}
\end{document}